\documentclass[runningheads,a4paper]{llncs}

\usepackage{amsmath,amssymb}
\usepackage{graphicx}
\usepackage{xcolor}

\usepackage{url}
\urldef{\mailsa}\path|{whu,myshao,cyang}@lbl.gov|
\urldef{\mailsb}\path|{andrea.cepellotti,jornada,sglouie}@berkeley.edu|
\urldef{\mailsc}\path|linlin@math.berkeley.edu|
\urldef{\mailsd}\path|kyle.thicke@duke.edu|

\newcommand{\Or}{\mathcal{O}}

\newcommand{\bpsi}{\bar{\psi}}

\newcommand{\hr}[1]{\hat{\vr}_{#1}}
\newcommand{\hbse}{H_\text{BSE}}
\newcommand*{\conj}[1]{\overline{#1}}
\renewcommand{\Im}{\mathrm{Im}}

\newcommand{\bvec}[1]{\mathbf{#1}}
\newcommand{\vr}{\bvec{r}}

\begin{document}

\mainmatter  

\title{Accelerating Optical Absorption Spectra and Exciton Energy Computation for Nanosystems via Interpolative Separable Density Fitting}

\titlerunning{Density fitting for GW and BSE calculations}

%
%
\author{Wei Hu$^1$
\and Meiyue Shao$^1$ \and Andrea Cepellotti$^2$ \and Felipe
H.~da~Jornada$^2$ \and Lin Lin$^{3, 1}$ \and Kyle Thicke$^{4}$ \and
Chao Yang$^1$ \and Steven G. Louie$^2$}
\authorrunning{Wei Hu et al.}

\institute{$^1$Computational Research Division,
Lawrence Berkeley National Laboratory,\\
Berkeley, California 94720, United States\\
\mailsa\\
$^2$Department of Physics,
University of California, Berkeley\\
Berkeley, California 94720, United States\\
\mailsb\\
$^3$Department of Mathematics,
University of California, Berkeley\\
Berkeley, California 94720, United States\\
\mailsc\\
$^4$Department of Mathematics,
Duke University,\\
Durham, NC 27708, United States\\
\mailsd}

%
%

\toctitle{}
\tocauthor{}
\maketitle

\begin{abstract}

We present an efficient way to solve the Bethe--Salpeter equation
(BSE), a model for the computation of absorption spectra in molecules and solids
that includes electron--hole excitations. Standard approaches
to construct and diagonalize the Bethe--Salpeter Hamiltonian require
at least $\Or(N_e^5)$ operations, where $N_e$ is proportional to the number
of electrons in the system, limiting its application to small systems.
Our approach is based on the interpolative separable density fitting (ISDF) technique to
construct low rank approximations to the bare and screened exchange
operators associated with the BSE Hamiltonian. This approach
reduces the complexity of the Hamiltonian construction to
$\Or(N_e^3)$ with a much smaller pre-constant. Here, we implement the
ISDF method for the BSE calculations within the Tamm--Dancoff
approximation (TDA) in the BerkeleyGW software package. We show that
ISDF-based BSE calculations in molecules and solids reproduce
accurate exciton energies and optical absorption spectra with
significantly reduced computational cost.


\end{abstract}

\section{Introduction} \label{sec:Introduction}

Many-Body Perturbation Theory~\cite{PR_46_618_1934_DMPT} is a
powerful tool to describe one-particle and two-particle excitations,
and to obtain excitation energies and absorption spectra in
molecules and solids~\cite{RMP_74_601_2002}. Within many-body
perturbation theory, Hedin's GW
approximation~\cite{PR_139_A796_1965_GW} has been successfully used
to compute quasi-particle (one-particle) excitation energies and the
Bethe--Salpeter equation (BSE)~\cite{PR_84_1232_1951_BSE} describes
the interaction of a electron--hole pair (two-particle excitation)
produced during the spectral absorption in molecules and
solids~\cite{PRB_62_4927_2000}. Good agreement between theory and
experiment can only be achieved taking into account such
electron--hole interaction by solving the BSE. The BSE is an
eigenvalue problem. In the context of optical absorption, the
eigenvalues of the Bethe--Salpeter Hamiltonian (BSH) are related to
quasi-particle exciton energies and the corresponding eigenfunctions
yield the exciton wavefunctions.

The BSH has a special block structure to be shown in the next
section. It consists of bare and screened exchange kernels that
depend on the products of single-particle orbitals obtained from a
mean-field calculation. The evaluation of these kernels requires at
least $\Or(N_e^5)$ operations in a conventional approach, which is
very costly for large complex systems that contain hundreds or even
thousands of atoms. There are several methods, which have been
developed recently to generate a reduced basis set, to reduce such
high computational cost for the BSE
calculations~\cite{JCP_334_221_2017,MolPhys_114_1148_2016,PRB_92_075422_2015,PRB_95_075415_2017,JCP_133_164109_2010}.

In this paper, we present a more efficient way to construct the BSH.
Such a construction allows the BSE to be solved efficiently by an
iterative diagonalization scheme. Our approach is based on the
recently developed interpolative separable density fitting (ISDF)
decomposition~\cite{JCP_302_329_2015}. This ISDF decomposition has
been applied to accelerate a number of applications in computational
chemistry and materials science, including two-electron integral
computation~\cite{JCP_302_329_2015}, correlation energy in the
random phase approximation~\cite{LuThicke2017}, density functional
perturbation theory~\cite{LinXuYing2017}, and hybrid density
functional calculations~\cite{JCTC_13_5420_2017_ISDF}. Such a
decomposition is used to approximate the matrix consisting of
products of single-particle orbital pairs as the product of a matrix
consisting of a small number of numerical auxiliary basis vectors
and an expansion coefficient matrix~\cite{JCTC_13_5420_2017_ISDF}.
This low rank approximation effectively allows us to construct low
rank approximations to the bare and screened exchange kernels. The
construction of ISDF compressed BSE Hamiltonian matrix only requires
$\Or(N_{e}^3)$ operations if the rank of the numerical auxiliary
basis can be kept at $\Or(N_{e})$ and if we keep the
bare and screened exchange kernel in a low rank factored form.
This is considerably more efficient than the $\Or(N_{e}^5)$
complexity required in a conventional approach.


By keeping these kernels in the decomposed form, the matrix--vector
multiplications required in iterative diagonalization procedures for
computing the desired eigenvalues and eigenvectors of $\hbse$ can be
performed efficiently. We may also use these efficient matrix--vector
multiplications in a structure preserving Lanczos
algorithm~\cite{arXiv_1611.02348_2016} to obtain an approximate
absorption spectrum, without explicitly diagonalizing the
approximate $\hbse$.

We have implemented the ISDF based BSH construction in the
BerkeleyGW software package~\cite{CPC_183_1269_2012_BerkeleyGW}. We
demonstrate that this approach can produce accurate exciton energies
and optical absorption spectra for molecules and solids, and reduce
the computational cost associated with the construction of the BSE
Hamiltonian significantly compared to the
algorithms used in the existing version of the BerkeleyGW software
package.

\section{Bethe--Salpeter equation}\label{sec:BSE}

The Bethe--Salpeter equation is an eigenvalue problem of the form
\begin{equation}
H_\text{BSE} X = X E, \label{eq:BSE}
\end{equation}
where $X$ is an exciton wavefunction, $E$ is the corresponding exciton
energy and $H_\text{BSE}$ is the Bethe--Salpeter Hamiltonian that
has the following block structure
\begin{equation}
   H_\text{BSE} =
  \begin{bmatrix}
   D + 2V_A - W_A &  2V_B - W_B \ \\
   -2\conj{V}_B + \conj{W}_B & - D -2\conj{V}_A + \conj{W}_A \ \\
  \end{bmatrix},
\label{eq:BSEHamiltonian}
\end{equation}
%
where $D(i_vi_c,j_vj_c) = (\epsilon_{i_c} -
\epsilon_{i_v})\delta_{i_vj_c}\delta_{i_cj_c}$ is an
$(N_vN_c)\times(N_vN_c)$ diagonal matrix with $\epsilon_{i_v}$,
$i_v=1,2,\dotsc,N_v$ being the quasi-particle energies associated
with valence bands and $\epsilon_{i_c}$,
$i_c=N_v+1,N_v+2,\dotsc,N_v+N_c$ being the quasi-particle energies
associated with conduction bands. These quasi-particle energies are
typically obtained from the so-called GW
calculation~\cite{PRB_62_4927_2000}. The $V_A$ and $V_B$ matrices
represent the (bare) {\em exchange} of electron--hole pairs, and the
$W_A$ and $W_B$ matrices are often referred to as the {\em direct}
terms that describe screened exchange of electron--hole pairs. These
matrices are defined as follows:
\begin{equation}
\begin{split}
V_A(i_vi_c,j_vj_c) &=\int \bpsi_{i_c}(\bvec{r})\psi_{i_v}(\bvec{r})
V(\bvec{r},\bvec{r'})\bpsi_{j_v}(\bvec{r'})\psi_{j_c}(\bvec{r'})\,\mathrm{d}\bvec{r}\,\mathrm{d}\bvec{r'},
\\
V_B(i_vi_c,j_vj_c) &=\int \bpsi_{i_c}(\bvec{r})\psi_{i_v}(\bvec{r})
V(\bvec{r},\bvec{r'})\bpsi_{j_c}(\bvec{r'})\psi_{j_v}(\bvec{r'})\,\mathrm{d}\bvec{r}\,\mathrm{d}\bvec{r'},
\\
W_A(i_vi_c,j_vj_c)
&=\int\bpsi_{i_c}(\bvec{r})\psi_{j_c}(\bvec{r})W(\bvec{r},\bvec{r'})
\bpsi_{j_v}(\bvec{r'})\psi_{i_v}(\bvec{r'}) \,\mathrm{d} \bvec{r}
\,\mathrm{d}\bvec{r'}, \\
 W_B(i_vi_c,j_vj_c)
&=\int\bpsi_{i_c}(\bvec{r})\psi_{j_v}(\bvec{r})W(\bvec{r},\bvec{r'})
\bpsi_{j_c}(\bvec{r'})\psi_{i_v}(\bvec{r'}) \,\mathrm{d} \bvec{r}
\,\mathrm{d}\bvec{r'},
\end{split}
\label{eq:BSEreal}
\end{equation}
where $\psi_{i_{v}}$ and $\psi_{i_{c}}$ are usually taken to be the
valence and conduction single-particle orbitals obtained from a
Kohn--Sham density functional theory (KSDFT) calculation
respectively, and $V(\bvec{r},\bvec{r'})$ and
$W(\bvec{r},\bvec{r'})$ are the bare and screened Coulomb operators.
Both $V_A$ and $W_A$ are Hermitian, whereas $V_B$ and $W_B$ are
complex symmetric in general. In the so-called Tamm--Dancoff
approximation (TDA)~\cite{RMP_74_601_2002}, both $V_B$ and $W_B$ are
neglected and set to zeros in Equation~\eqref{eq:BSEHamiltonian}. As
a result, $\hbse$ is Hermitian in this case, and we only need to be
concerned with the upper left block of $H_\text{BSE}$.

Let $M_{cc}(\bvec{r}) = \{\psi_{i_c}\bpsi_{j_c}\}$,
$M_{vc}(\bvec{r}) = \{\psi_{i_c}\bpsi_{i_v}\}$, and
$M_{vv}(\bvec{r}) = \{\psi_{i_v}\bpsi_{j_v}\}$ be matrices that
consist of products of discretized orbital pairs in real space, and
$\hat{M}_{cc}(\bvec{G})$, $\hat{M}_{vc}(\bvec{G})$,
$\hat{M}_{vv}(\bvec{G})$ be the reciprocal space representation of
these matrices. Equations~\eqref{eq:BSEreal} can then be succinctly
written as
%
%
\begin{equation}
\begin{split}
V_A &= \hat{M}_{vc}^{*} \hat{V} \hat{M}_{vc}, \\
W_A &= \mbox{reshape}(\hat{M}_{cc}^{*} \hat{W} \hat{M}_{vv}),
\end{split}
\label{eq:BSEreciprocal}
\end{equation}
where $\hat{V}$ and $\hat{W}$ are reciprocal space representations
of the bare and screened exchange operators $V$ and $W$,
respectively, and the reshape function is used to map the
$(i_cj_c,i_vj_v)$th element on the right-hand side of
\eqref{eq:BSEreciprocal} to the $(i_ci_v,j_cj_v)$th element of
$W_A$.  A similar set of equations can be derived for $V_B$
and $W_B$. However, in this paper, we will mainly focus on the TDA
model.

The reason that the right-hand sides
of~\eqref{eq:BSEreciprocal} are computed in the reciprocal space is
that $\hat{V}$ is diagonal and an energy cutoff is often adopted to limit
the number of the Fourier components used to represent $\psi_i$. As
a result, the leading dimension of $\hat{M}_{cc}$, $\hat{M}_{vc}$ and
$\hat{M}_{cc}$, which we denote by $N_g$, is often much smaller than
that of $M_{cc}$, $M_{vc}$ and $M_{vv}$, which we denote by $N_r$.
%

In additional to performing $\Or(N_e^2)$ Fast Fourier transforms
(FFTs) to obtain $\hat{M}_{cc}$, $\hat{M}_{vc}$ and $\hat{M}_{vv}$
from $M_{cc}$, $M_{vc}$ and $M_{vv}$, respectively, we need to
perform at least
\begin{equation}
\Or(N_gN_c^2N_v^2 + N_g^2N_cN_v)
\label{eq:cost}
\end{equation}
floating-point operations to obtain $V_A$ and $W_A$ using
matrix--matrix multiplication operations.

Note that, to achieve high accuracy with a large basis set, such as
the plane wave basis set, $N_g$ is typically much larger than $N_c$
or $N_v$. The number of occupied bands is either $N_{e}$ or $N_e/2$
depending on how spin is counted. In many existing calculations,
the actual number of conducting band $N_{c}$ included in the
calculation is often a small multiple of $N_{v}$, whereas $N_g$
is often 100--10000 $\times N_e$ ($N_r \sim 10 \times N_g$). Therefore, the
second term in Equation~\eqref{eq:cost}, which accounts for the cost
of multiplying $\hat{W}_A$ with $\hat{Z}$ in
Equation~\eqref{eq:BSEreciprocal} can be much larger than other
parts under the Tamm--Dancoff approximation (TDA) in the BSE
calculations.

\section{Interpolative separable density fitting (ISDF) decomposition}\label{sec:isdf}

In order to reduce the computational complexity, we seek to minimize
the number of integrals we need to perform in
Equation~\eqref{eq:BSEreal}. This is possible if we can rewrite the
matrix $M_{ij}$, where the labels $i$ and $j$ are indices of
either valence or conducting orbitals,
as the product of a matrix $\Theta_{ij}$ that contains a set of
$N_{ij}^t$ linearly independent auxiliary basis vectors with
$N_{ij}^t \approx tN_e \ll \Or(N_e^2)$ ($t$ is a small constant
referred as a rank truncation
parameter)~\cite{JCTC_13_5420_2017_ISDF} and an expansion
coefficient matrix $C_{ij}$. For large problems, the number of
columns of $M_{ij}$, which is either $\Or(N_vN_c)$, $\Or(N_v^2)$, or
$\Or(N_c^2)$, is typically larger than the number of grid points
$N_r$ on which $\psi_n(\bvec{r})$ is sampled, i.e., the number of
rows in $M_{ij}$. As a result, $N_{ij}^t$ should be much smaller than the
number of columns of $M_{ij}$. Even when a cutoff is used to limit
the size of $N_c$ or $N_v$ so that the number of columns in $M_{ij}$
is much less than $N_g$, we can still approximate $M_{ij}$ by
$\Theta_{ij} C_{ij}$ with a $\Theta_{ij}$ that has a smaller rank
$N_{ij}^t \sim t\sqrt{N_iN_j}$.
%

To simplify our discussion, let us drop the subscript of $M$,
$\Theta$ and $C$ for the moment, and describe the basic idea of
ISDF. The optimal low rank approximation of $M$ can be obtained from
a singular value decomposition. However, the complexity of this
decomposition is at least $\Or(N_r^2 N_e^2)$ or $\Or(N_e^4)$. An
alternative decomposition, which is close to optimal, but has a much
favorable complexity has recently been developed. This type of
decomposition is called interpolative separable density fitting
(ISDF)~\cite{JCTC_13_5420_2017_ISDF}, which we will now describe.

In ISDF, instead of computing $\Theta$ and $C$ simultaneously,
we fix the coefficient matrix $C$ first, and
determine the auxiliary basis matrix $\Theta$ by solving a linear
least squares problem
\begin{equation}
  \min \| M - \Theta C \|_F^2,
  \label{eqn:isdflineq}
\end{equation}
where each column of $M$ is given by $\psi_{i}(\vr)\bpsi_{j}(\vr)$
sampled on a dense real space grids $\{\vr_{i}\}_{i=1}^{N_r}$, and
$\Theta = [\zeta_1, \zeta_2, \dotsc, \zeta_{N^t}]$ contains the
auxiliary basis vectors to be determined, $\|\cdot \|_F$ denotes the
Frobenius norm.

We choose $C$ to be a matrix that consists of
$\psi_{i}(\vr)\bpsi_{j}(\vr)$ evaluated at a subset of $N^t$
carefully chosen real space grid points, with $N^t \ll N_r$ and $N^t
\ll N_e^2$, i.e., each column of $C$ indexed by $(i,j)$ is given by
\[ [ \psi_{i}(\hr{1})\bpsi_{j}(\hr{1}), \cdots,
    \psi_{i}(\hr{k})\bpsi_{j}(\hr{k}), \cdots,
    \psi_{i}(\hr{N^t})\bpsi_{j}(\hr{N^t})]^{\mathsf{T}}.
\]

If the minimum of the objective function in Equation~\eqref{eqn:isdflineq} is
zero, the product of $\Theta$ and a column of $C$ can be viewed as an
interpolation of corresponding function
$\{\psi_{i}(\bvec{r})\bpsi_{j}(\bvec{r})\}$ in $M$. However, in
general, we cannot expect the minimum of \eqref{eqn:isdflineq} be
zero. The least squares minimizer is given by
\begin{equation}
\Theta = MC^* (CC^*)^{-1}. \label{eq:Theta}
\end{equation}

It may appear that the matrix--matrix multiplications $MC^*$ and
$CC^*$ take $\Or(N_{e}^{4})$ operations because the size of $M$ is
$N_r \times \Or(N_{e}^2)$ and the size of $C$ is $N^t \times
\Or(N_{e}^2)$. However, both multiplications can be carried out with
fewer operations due to the separable structure of $M$ and
$C$~\cite{JCTC_13_5420_2017_ISDF}. As a result, the computational
complexity for computing the interpolation vectors is
$\Or(N_{e}^{3})$.

Intuitively, the least squares problem in
Equation~\eqref{eqn:isdflineq} is easier to solve when the rows of
$C$, which can be selected from the rows of $M$, are maximally
linearly independent. This task can be achieved by performing a QR
factorization of $M^{\mathsf{T}}$ with column pivoting
(QRCP)~\cite{SIAM_13_727_1992_QRCP}. In QRCP, we choose a
permutation $\Pi$ such that the factorization
\begin{equation}
  M^{\mathsf{T}} \Pi = QR
  \label{eqn:QRCP}
\end{equation}
yields a unitary matrix $Q$ and an upper triangular matrix $R$ with
decreasing matrix elements along the diagonal of $R$. The magnitude
of each diagonal element $R$ indicates how important the
corresponding column of the permuted $M^{\mathsf{T}}$ is, and whether the
corresponding grid point should be chosen as an interpolation point.
The QRCP decomposition can be terminated when the $(N^t+1)$-st
diagonal element of $R$ becomes less than a predetermined threshold.
The leading $N^t$ columns of the permuted $M^{\mathsf{T}}$ are considered to be
maximally linearly independent numerically. The corresponding grid
points are chosen as the interpolation points. The indices for the
chosen interpolation points $\hr{N^t}$ can be obtained from indices
of the nonzero entries of the first $N^t$ columns of the permutation
matrix~$\Pi$.

Roughly speaking, the QRCP moves matrix columns
of $M^{\mathsf{T}}$ with large norms to the left, and pushes matrix columns
of $M^{\mathsf{T}}$ with small norms to the right. Note that the square of the
vector 2-norm of the column of $M^{T}$ associated with $\vr$ is just
\begin{equation}
  \sum_{i,j=1}^{N} \bigl|\psi_{i}(\vr) \bpsi_{j}(\vr)\bigr|^2 =
  \biggl(\sum_{i=1}^{N} \bigl|\psi_{i}(\vr)\bigr|^2\biggr)
  \biggl(\sum_{j=1}^{N} \bigl|\psi_{j}(\vr)\bigr|^2\biggr).
  \label{}
\end{equation}
In the case when $\{\psi_{i}\}$ and $\{\psi_{j}\}$ both belong to the
set of occupied orbitals, the norm of each column of $M^{\mathsf{T}}$ is
simply the electron density. Hence the interpolation points chosen
by QRCP tend to locate in areas where the electron density is
relatively large. Once a column is selected, all other columns are
immediately orthogonalized with respect to the chosen column. Hence
nearly linearly dependent matrix columns will not be selected
repeatedly. As a result, the interpolation points chosen by QRCP are
well separated spatially. Notice that the standard QRCP procedure has
a high computational cost of $\Or(N_{e}^2N_{r}^2) \sim \Or(N_{e}^4)$. But
it can be combined with the randomized sampling
method~\cite{JCP_302_329_2015} so that its cost is reduced to
$\Or(N_{r}N_{e}^2) \sim \Or(N_{e}^3)$.

\section{Low rank representations of bare and screened exchange operators via ISDF}\label{sec:isdfbse}


Applying the ISDF decomposition to $M_{cc}$, $M_{vc}$ and $M_{vv}$
yields
\begin{equation}
\begin{split}
M_{cc} &\approx \Theta_{cc} C_{cc},  \\
M_{vc} &\approx \Theta_{vc} C_{vc}, \\
M_{vv} &\approx \Theta_{vv} C_{vv}.
\end{split}
\label{eq:ISDFpairs}
\end{equation}
It follows from Equations~\eqref{eq:BSEreal}, \eqref{eq:BSEreciprocal}
and \eqref{eq:ISDFpairs} that the exchange and direct terms of the
BSE Hamiltonian can be written as
\begin{equation}
\begin{split}
V_A  &= C_{vc}^{*} \widetilde{V}_A C_{vc}, \\
W_A  &= \mbox{reshape}(C_{cc}^{*} \widetilde{W}_A C_{vv}), \\
\end{split}
\label{eq:vawaisdf}
\end{equation}
where $\widetilde{V}_A=\hat{\Theta}_{vc}^{*}
\hat{V} \hat{\Theta}_{vc}$ and
$\widetilde{W}_A = \hat{\Theta}_{cc}^{*}
\hat{W} \hat{\Theta}_{vv}$ are the \emph{projected}
exchange and direct terms under the auxiliary
basis $\hat{\Theta}_{vc}$, $\hat{\Theta}_{cc}$ and
$\hat{\Theta}_{vv}$. Here, $\hat{\Theta}_{vc}$, $\hat{\Theta}_{cc}$
and $\hat{\Theta}_{vv}$ are reciprocal space representations
of $\Theta_{vc}$, $\Theta_{cc}$ and $\Theta_{vv}$, respectively, that
can be obtained via FFTs,

Note that the dimension of the matrix $C_{cc}^{*} \widetilde{W}_A
C_{cc}$ on the right-hand side of Equation~\eqref{eq:vawaisdf} is
$N_c^2 \times N_v^2$.  It needs to be reshaped into a matrix of
dimension $N_v N_c \times N_vN_c$ according to the mapping
$W_A(i_cj_c,i_vj_v) \rightarrow W_A(i_vi_c,j_vj_c)$ before it can
combined with $V_A$ matrix to construct the BSH.

Once the ISDF approximations for $M_{vc}$, $M_{cc}$ and
$M_{vv}$ are available, the cost for constructing a
low rank approximation to the exchange and direct terms reduced to
that of computing the projected exchange and direct kernels
$\hat{\Theta}_{vc}^{*} \hat{V} \hat{\Theta}_{vc}$ and
$\hat{\Theta}_{cc}^{*} \hat{W} \hat{\Theta}_{vv}$,
respectively.  If the ranks of $\Theta_{vc}$, $\Theta_{cc}$
and $\Theta_{vv}$ are $N_{vc}^t$, $N_{cc}^t$ and $N_{vv}^t$,
respectively, then the computational complexity for computing the
compressed exchange and direct kernels is
$\Or(N_{vc}^t N_{vc}^t N_g + N_{cc}^tN_{vv}^t N_g + N_{vv}^t N_g^2)$, which
is significantly lower than the complexity of the conventional
approach given in \eqref{eq:cost}.  When $N_{vc}^t \sim t\sqrt{N_vN_c}$, $N_{cc}^t \sim t\sqrt{N_cN_c}$ and
$N_{vv}^t \sim t\sqrt{N_vN_v}$ are on the order of $N_e$, the complexity of constructing
the compressed kernels is $\Or(N_e^3)$.

\section{Iterative diagonalization of the BSE Hamiltonian}

In the conventional approach, exciton energies and wavefunctions
can be computed by using the recently developed BSEPACK
library~\cite{LinearAlgebraAppl_488_148_2016,BSEPACK_UserGuide_2016}
to diagonalize the BSE Hamiltonian $H_\text{BSE}$. When TDA
is adopted, we may also just use a standard diagonalization
procedure implemented in the ScaLAPACK~\cite{ScaLAPACK_2011}
library.

When ISDF is used to construct low rank approximations to
the bare and screened exchange operators $V_A$ and $W_A$, we
should keep both matrices in the factored form given
by Equation~\eqref{eq:vawaisdf}. This is because that multiplying the
matrices on the right-hand sides of Equation~\eqref{eq:vawaisdf} would
require at least $\Or(N_e^5)$ operations, which is higher than the cost
for using the ISDF procedure to construct low rank
approximations to the BSH. Instead of using BSEPACK or ScaLAPACK
which has a complexity of $\Or(N_e^6)$, we propose to use
iterative methods to diagonalize the approximate BSH
constructed via the ISDF decomposition.


When TDA is used, several iterative methods such as the
Lanczos~\cite{JRNBS_45_255_1950_Lanczos} and LOBPCG~\cite{SIAMJSC_23_517_2001_LOBPCG} algorithms can be used to compute a
few desired eigenvalues of the $H_\text{BSE}$. In each step of these
algorithms, we need to multiply $H_\text{BSE}$ with a vector $x$ of
size $N_vN_c$. When $V_A$ is kept in the factored form given by
\eqref{eq:vawaisdf}, $V_Ax$ can be evaluated as three matrix vector
multiplications performed in sequence, i.e.,
\begin{equation}
V_A x \leftarrow C_{vc}^* \bigl[\widetilde{V_A} (C_{vc} x)\bigr].
\label{eq:vamv}
\end{equation}
The complexity of these calculations is $\Or(N_vN_c N_{vc}^t)$.
If $N_{vc}^t$ is on the order of $N_e$, then each $V_A x$ can be
carried out in $\Or(N_e^3)$ operations.

Because $C_{cc}^* \widetilde{W_A} C_{vv}$ cannot be multiplied
with a vector $x$ of size $N_vN_c$ before it is reshaped,
a different multiplication scheme needs to be used. It follows
from the separable nature of $C_{vv}$ and $C_{cc}$ that this
multiplication can be succinctly described as
\begin{equation}
W_A x = \text{reshape}\left[ \Psi_c^{*} \bigl(\widetilde{W} \odot
(\Psi_{c} X \Psi_{v}^*)\bigr)\Psi_v
\right],
\label{eq:wamv}
\end{equation}
where $X$ is a $N_c \times N_v$ matrix reshaped from the vector $x$,
$\Psi_{c}$ is a $N_{cc}^t \times N_c$ matrix containing
$\psi_{i_c}(\hat{r}_k)$ as its elements, $\Psi_{v}$ is a $N_{vv}^t \times N_v$
matrix containing $\psi_{i_v}(\hat{r}_k)$ as its elements, and
$\odot$ denotes componentwise multiplication (Hadamard product).
The reshape function is used to turn the $N_c\times N_v$ matrix--matrix
product back into a size $N_vN_c$ vector. If $N_{vv}^t$ and $N_{cc}^t$
are on the order of $N_e$, then all matrix--matrix multiplications
in Equation~\eqref{eq:wamv} can be carried out in $\Or(N_e^3)$ operations.
This make the complexity of each step of the iterative method
$\Or(N_e^3)$. If the number of iterative steps required to reach
convergence is not excessively large, then the ISDF enabled
iterative diagonalization can be carried out in $\Or(N_e^3)$ operations.

\section{Estimating optical absorption spectra without diagonalization}
If we have all eigenpairs of $H_\text{BSE}$, we can easily obtain
the optical absorption spectrum, which is the imaginary part of the
dielectric function defined as
\begin{equation}
\varepsilon_2(\omega) = \Im\biggl[
\frac{8\pi^2e^2}{\Omega}
d_r^H\bigl((\omega-\mathrm{i}\eta){I}-H_\text{BSE}\bigr)^{-1}d_l
\biggr],
\label{eq:absorption}
\end{equation}
where $\Omega$ is the volume of the primitive cell, $e$ is the
elementary charge, $d_r$ and $d_l$ are the right and left optical
transition vectors, and $\eta$ is a broadening factor used to
account for the lifetime of excitation.  However, it can become
prohibitively expensive to use an iterative diagonalization method
to compute all eigenpairs of $H_\text{BSE}$.

Fortunately, it is possible to use a structure preserving iterative
method to estimate the optical absorption spectrum without explicitly
computing all eigenpairs of $H_\text{BSE}$.
In Ref.~\cite{JCTC_11_5197_2015,arXiv_1611.02348_2016}, we developed
a structure preserving Lanczos algorithm for estimating
the optical spectrum. The algorithm has been implemented in the
BSEPACK~\cite{BSEPACK_UserGuide_2016} library. When TDA is used,
the standard Lanczos algorithm can be used to estimate
the absorption spectrum.  When our objective is
to obtain the basic shape of the absorption spectrum to identify
where the major peaks are, it is not necessary to compute
all eigenpairs of $H_\text{BSE}$. As a result, the accuracy
required to construct approximate bare and screened exchange operators
used in BSE can possibly be lowered, thereby allowing us to
use a more aggressive truncation threshold in ISDF to further reduce
the cost of $H_\text{BSE}$ construction. We will demonstrate
this possibility in the next section.

\section{Numerical results} \label{sec:Result}

In this section, we demonstrate the accuracy and efficiency of the ISDF
method when it is used to compute exciton energies and optical
absorption spectrum in the BSE framework.  We implemented the ISDF based BSH
construction in the BerkeleyGW software package~\cite{CPC_183_1269_2012_BerkeleyGW}.
BerkeleyGW is a massively parallel computational package that uses
a many-body perturbation theory and Green's function formalism
to study quasi-particle excitation energies and optical absorption of
nanosystems. We use the \emph{ab initio} software package
Quantum ESPRESSO (QE)~\cite{JPCM_21_395502_2009_QE} to compute the ground-state
quantities required in the GW and BSE calculations. In our Quantum ESPRESSO
density functional theory (DFT) based electronic structure calculations, we
use Hartwigsen--Goedecker--Hutter (HGH) norm-conserving
pseudopotentials~\cite{PRB_58_3641_1998_HGH} and the
LDA~\cite{PRB_54_1703_1996_LDA} exchange--correlation functional.
All the calculations were carried out on a single core at the Cori%
\footnote{https://www.nersc.gov/systems/cori/}
systems at the National Energy Research Scientific Computing Center
(NERSC).

\begin{figure}[!tb]
\centering
\includegraphics[width=0.9\textwidth]{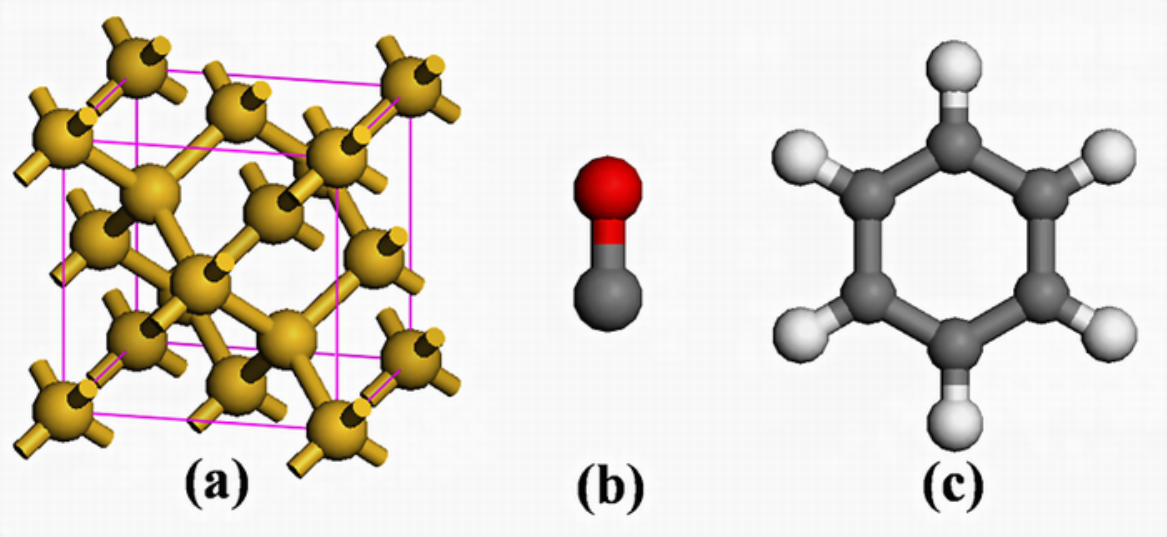}
\caption{Atomic structures of (a) a bulk silicon Si$_{8}$ unit cell,
(b) carbon monoxide (CO) and (c) benzene (C$_6$H$_6$) molecules.
The white, gray, red, and yellow balls denote hydrogen, carbon, oxygen,
and silicon atoms, respectively.} \label{fig:Structure}
\end{figure}

We performed calculations for three systems. They consist of a bulk
silicon Si$_{8}$ system and two molecular systems: carbon monoxide
(CO) and benzene (C$_6$H$_6$) as plotted in
Fig.~\ref{fig:Structure}. All systems are closed shell systems, and
the number of occupied bands is $N_v = N_{e}/2$, where $N_e$ is the
valence electrons in the system.

\subsection{Accuracy}\label{sec:Accuracy}

We first measure the accuracy of the ISDF method by comparing
the computed eigenvalues of the BSH matrices constructed with and without the ISDF
decomposition.

In our test, we set the plane wave energy cutoff required in the QE
calculations to $E_\text{cut} = 10$ Ha, which is relatively low.
However, this is sufficient for assessing the effect of ISDF. Such a
choice of $E_\text{cut}$ results in $N_r = 35937$ and $N_g = 2301$
for the Si$_{8}$ system, $N_r = 19683$ and $N_g = 1237$ for the CO
molecule ($N_v = 5$), $N_r = 91125$ and $N_g = 6235$ for the benzene
molecule. We only include the lowest $N_c$ conducting bands in the
BSE calculation. The number of active conduction bands ($N_c$) and
valence bands ($N_v$), the number of reciprocal grids and the
dimensions of the corresponding BSE Hamiltonian $H_\text{BSE}$ for
these three systems are listed in Table~\ref{Structure}.

\begin{table}[!tb]
\centering \caption{Parameters of system size for bulk silicon
Si$_{8}$, carbon monoxide (CO) and  benzene (C$_6$H$_6$) molecules
used for constructing corresponding BSE Hamiltonian
$H_{BSE}$.} \label{Structure}
\begin{tabular}{ccccccccccc} \ \\
\hline \hline
System    && $N_r$ && $N_g$ && $N_v$ && $N_c$ && dim($H_{BSE}$) \ \\
 \hline
  Si$_8$  && 35937 && 2301  && 16    &&  64   &&  2048 \ \\
  CO      && 19683 && 1237  &&  5    &&  60   &&  600  \ \\
  Benzene && 91125 && 6235  && 15    &&  60   &&  1800 \ \\
\hline \hline
\end{tabular}
\end{table}

\begin{figure}[!tb]
\centering
\includegraphics[width=0.95\textwidth]{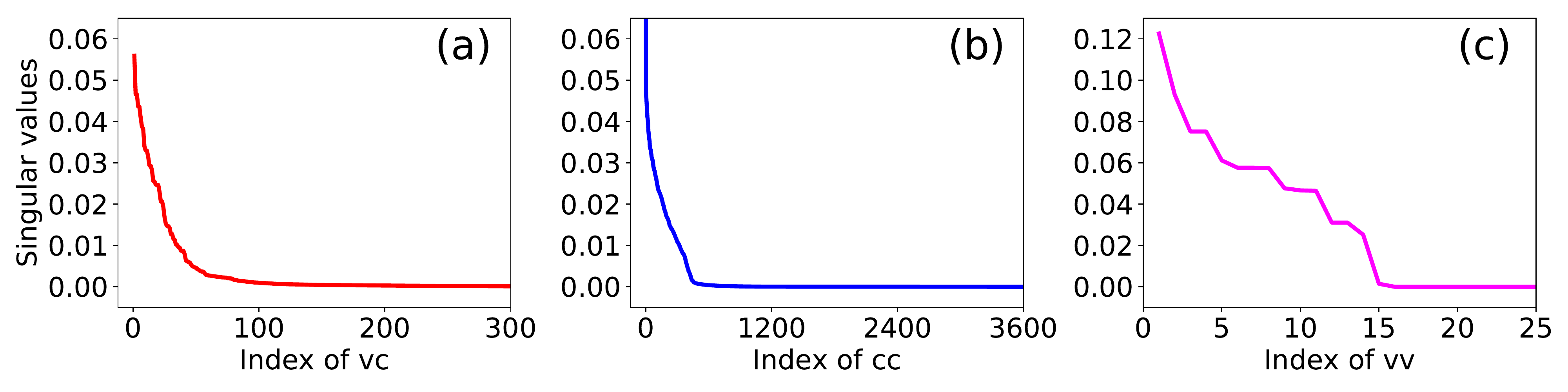}
\caption{The singular values of (a) $M_{vc}(\bvec{r}) =
\{\psi_{i_c}(\bvec{r})\bpsi_{i_v}(\bvec{r})\}$ ($N_{vc} = 300$), (b)
$M_{cc}(\bvec{r}) = \{\psi_{i_c}(\bvec{r})\bpsi_{j_c}(\bvec{r})\}$
($N_{cc} = 3600$) and (c) $M_{vv}(\bvec{r}) =
\{\psi_{i_v}(\bvec{r})\bpsi_{j_v}(\bvec{r})\}$ ($N_{vv} = 25$).}
 \label{fig:COSVD}
\end{figure}

In Fig.~\ref{fig:COSVD}, we plot the singular values of the
matrices $M_{vc}(\bvec{r}) = \{\psi_{i_c}(\bvec{r})\bpsi_{i_v}(\bvec{r})\}$,
$M_{cc}(\bvec{r}) = \{ \psi_{i_c}(\bvec{r})\bpsi_{j_c}(\bvec{r})\}$
and $M_{vv}(\bvec{r}) =
\{\psi_{i_v}(\bvec{r})\bpsi_{j_v}(\bvec{r})\}$ associated with the
CO molecule. We observe that the singular values of these matrices
decay rapidly. For example, the leading 500 (out of 3600) singular
values of $M_{cc}(\bvec{r})$ decreases rapidly towards zero. All
other singular values are below $10^{-4}$. Therefore, the
numerical rank $N_{cc}^t$ of $M_{cc}$ is roughly 500 ($t$ = 8.3), or roughly 15\% of the
number of columns in $M_{cc}$. Consequently, we expect that the rank
of $\Theta_{cc}$ produced in ISDF decomposition can be set to 15\%
of $N_c^2$ without sacrificing the accuracy of the computed
eigenvalues.

\begin{figure}[!tb]
\centering
\includegraphics[width=0.6\textwidth]{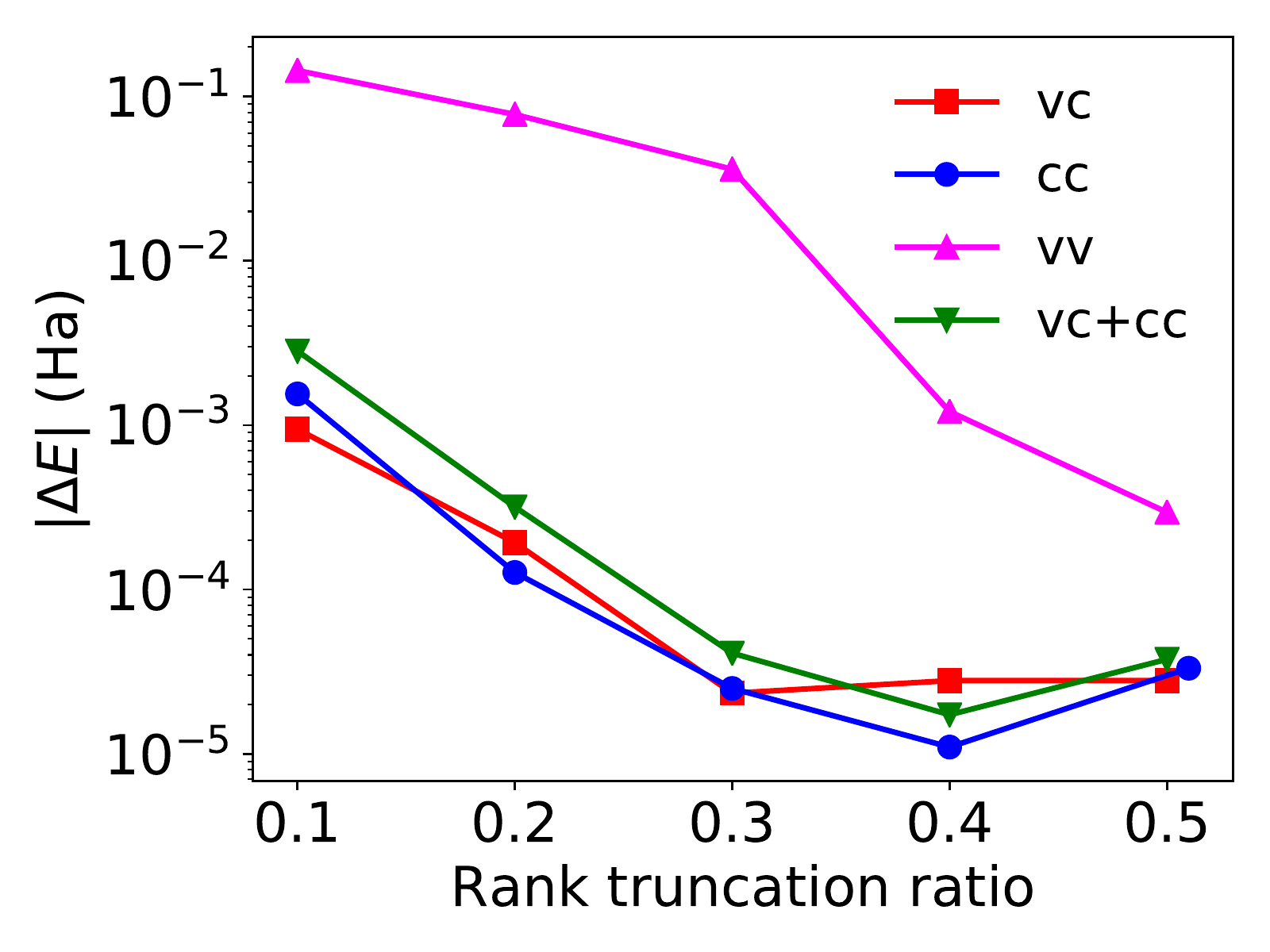}
\caption{The change of absolute error $\Delta E$ in the smallest
eigenvalue of $H_\text{BSE}$ associated with the Si$_8$ system with
respect to different truncation levels used in ISDF approximation of
$M_{vc}$, $M_{cc}$ and $M_{vv}$. The curves labeled by `vc', `cc',
`vv' correspond to calculations in which only one of the ratios
$N_{vc}^t/N_{vc}$, $N_{cc}^t/N_{cc}$ and $N_{vv}^t/N_{vv}$ changes
while all other parameters are held constant.  The curve labeled by
`vc + cc' corresponds to the calculation in which both
$N_{vc}^t/N_{vc}$ and $N_{cc}^t/N_{cc}$ change at the same rate ($N_{vv}^t = N_{vv}$).} \label{fig:Si8EigErr}
\end{figure}

This prediction is confirmed in Fig.~\ref{fig:Si8EigErr}, where
we plot the absolute difference between the lowest the exciton
energy of Si$_{8}$ computed with and without using ISDF to construct
$H_{\text{BSE}}$. To be specific, the error in the desired
eigenvalue is computed as $\Delta{E} = E_\text{ISDF} -
E_\text{BGW}$, where $E_\text{ISDF}$ is computed from the
$H_{\rm{BSE}}$ constructed with ISDF approximation, and
$E_\text{BGW}$ is computed from a standard $H_{\text{BSE}}$
constructed without using ISDF. We first vary one of the ratios
$N^{t}_{cc}/N_{cc}$, $N^t_{vc}/N_{vc}$ and $N^t_{vv}/N_{vv}$ while
holding the others at a constant of~$1$. We observe that the error in
the lowest exciton energy (positive eigenvalue) is around $10^{-3}$
Ha, when either $N^{t}_{cc}/N_{cc}$ or $N^t_{vc}/N_{vc}$ is set to
0.1 while the other ratios are held at 1. However, reducing
$N^t_{vv}/N_{vv}$ to 0.1 introduces a significant amount of error
(0.1 Ha) in the lowest exciton energy. This is likely due to the
fact that $N_v=16$ is too small. We then hold $N^t_{vv}/N_{vv}$ at
0.5 and let both $N^{t}_{cc}/N_{cc}$ and $N^t_{vc}/N_{vc}$ vary. The
variation of $\Delta{E}$ with respect to these ratios is also
plotted as in Fig.~\ref{fig:Si8EigErr}. We observe that the error in
the lowest exciton energy is still around $10^{-3}$~Ha even when
both $N^{t}_{cc}/N_{cc}$ and $N^t_{vc}/N_{vc}$ are set to 0.1.

%

\begin{figure}[!tb]
\centering
\includegraphics[width=0.95\textwidth]{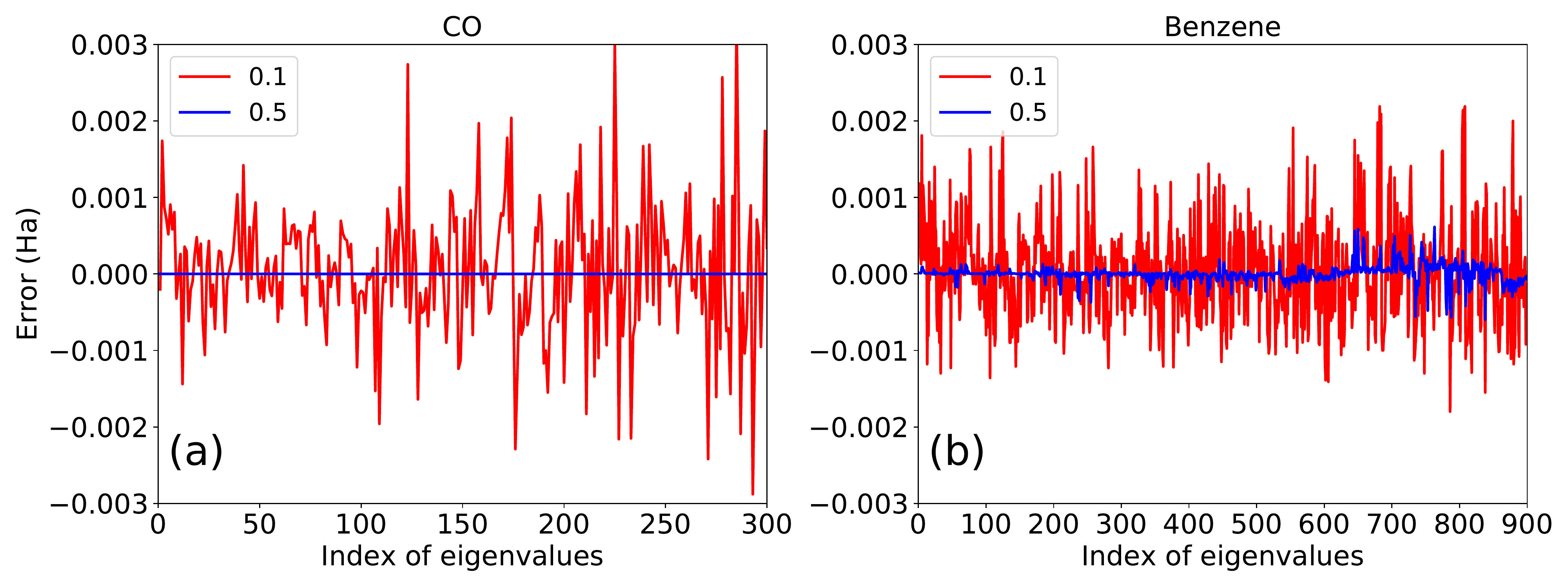}
\caption{Error in all eigenvalues of the BSH associated with
the CO (a) and benzene (b) molecules.
Two rank truncation ratios $N^t_{cc}/N_{cc} = 0.5$ ($t = 30.0$)
and $N^t_{cc}/N_{cc} = 0.1$ ($t = 6.0$) are used in the tests.}
\label{fig:COBenzeneEigenvalues}
\end{figure}

We then check the absolute error $\Delta{E}$ (Ha) of all the exciton
energies computed with the ISDF method by comparing them with the ones obtained from a conventional BSE calculation implemented in BerkeleyGW
for the CO and benzene molecules. As we can see from  Fig.~\ref{fig:COBenzeneEigenvalues}, the errors associated with these eigenvalues are all below $0.002$~Ha when $N^t_{cc}/N_{cc}$ is $0.1$.

\subsection{Efficiency}\label{sec:Efficiency}

At the moment, we have only implemented a sequential version of the
ISDF method within the BerkeleyGW software package. Therefore, our
efficiency test is limited in the size of the problem as well as the
number of conducting bands ($N_c$) we can include in the bare and
screened exchange operators. As a result, our performance
measurement does not fully reflect the computational complexity
analysis presented in the previous sections. In particular, taking
benzene as an example, $N_g = 6235$ is much larger than $N_v = 15$
and $N_c = 60 $, therefore the computational cost of $N_g^2N_v^2
\sim \Or(N_e^4)$ term is much higher than the $N_gN_v^2N_c^2 \sim
\Or(N_e^5)$ term in the conventional BSE calculations.


Nonetheless, in this section, we will demonstrate the benefit of
using ISDF to reduce the cost for constructing the BSE Hamiltonian
$H_\text{BSE}$. In Table~\ref{tab:isdf_time}, we focus on the
benzene example and report the wall-clock time required to construct
the ISDF approximations of the $M_{vc}$,
$M_{cc}$, and $M_{vv}$ matrices at different rank truncation
levels. Without using ISDF, it takes 746.0 seconds to
construct the reciprocal space representations of $M_{vc}$,
$M_{cc}$, and $M_{vv}$ in BerkeleyGW. Most of the timing is spent
in the large number of FFTs applied to $M_{vc}$,
$M_{cc}$, and $M_{vv}$ to obtain the reciprocal space representation of these matrices. We can clearly see that if only $N^t_{cc}/N_{cc}$ is changed from 0.5 ($t$ = 30.0) to 0.1 ($t$ = 6.0), the wall-clock time used to construct a low rank approximation to $M_{cc}$ can be reduced from 578.9 to 34.3 seconds. Furthermore, the total cost of computing $M_{vc}$, $M_{cc}$ and $M_{vv}$ can be further
reduced to around 1/19th that in a conventional approach
(39.3 vs.\ 746.0 seconds) if $N^{t}_{vc}/N_{vc}$,
$N^{t}_{vv}/N_{vv}$ and $N^{t}_{cc}/N_{cc}$ are all set to
0.1.


%
%
\begin{table}[!tb]
\centering \caption{The variation of time required to carry
out the ISDF decomposition of $M_{vc}$, $M_{vv}$ and $M_{cc}$
with respect to rank truncation ratio.}\label{tab:isdf_time}
\begin{tabular}{ccccccc} \ \\
\hline \hline
\multicolumn{3}{c}{Rank truncation ratio} & & \multicolumn{3}{c}{Time (s) for $M_{ij}(\bvec{r})$ }  \ \\
\hline
 $N_{vc}^{t}$/$N_{vc}$  &  $N_{vv}^{t}$/$N_{vv}$  & $N_{cc}^{t}$/$N_{cc}$ & & $M_{vc}$ & $M_{vv}$ & $M_{cc}$\ \\
\hline
1.0 & 0.5 & 0.5 && 157.0 &  5.8 & 578.9 \ \\
1.0 & 0.5 & 0.1 && 157.0 &  5.8 & 34.3 \ \\
0.1 & 0.1 & 0.1 && 4.3   &  0.7 & 34.3 \ \\
\hline \hline
\end{tabular}
\end{table}


Note that because ISDF decomposition is carried out on a real space
grid, its measured cost reflect the cost for performing
QRCP in real space. Even though QRCP with random sampling has
$\Or(N_e^3)$ complexity, it has a relatively large pre-constant
compared to the size of the problem. Hence the measured
cost of QRCP is relatively high in this case. Recently,
Dong et al.\ proposed a new
approach to find the interpolation points based on the centroidal
Voronoi tessellation (CVT) method~\cite{arXiv_2017_CVT}, which
offers a much less expensive alternative to the QRCP procedure when
the ISDF method is used in hybrid functional calculations. We will
explore this CVT method in the BSE-ISDF calculations in the future.

In Table~\ref{tab:bsh_time}, we report the wall-clock time required
to construct the projected bare and screened exchange matrices
$\widetilde{V}_A$ and $\widetilde{W}_A$ that appear in
Equation~\eqref{eq:vawaisdf} once the ISDF approximations of
$M_{vc}$, $M_{vv}$, and $M_{cc}$ become available.
Without ISDF, it takes
$1.574 + 4.198 = 5.772$ seconds to construct both $W_A$ and $V_A$.
When $N_{cc}^{t}$/$N_{cc}$ is set to 0.1 only,
the construction cost for $W_A$, which is the dominant cost, is reduced by
a factor of 2.8. Furthermore, if $N^t_{vc}/N_{vc}$, $N^t_{vv}/N_{vv}$
and $N^t_{cc}/N_{cc}$ are all set to 0.1. We reduce the
cost for constructing $\widetilde{V}_A$ and $\widetilde{W}_A$ by
a factor of 63.0 and 10.1, respectively.
Note that the original implementation of the $W_A$ and $V_A$ in BerkeleyGW
is much slower because the elements of $W_A$ and $V_A$ are integrated one by
one. For benzene, it take takes $103{,}154$ seconds ($28.65$ hours) to construct the BSE
Hamiltonian $H_\text{BSE}$ in the original BerkeleyGW code.

\begin{table}[!tb]
\centering \caption{The variation of time required to construct the
projected bare and screened exchange matrices $\widetilde{V}_A$ and
$\widetilde{W}_A$ exhibited by the ISDF method respect to rank
truncation ratio.}\label{tab:bsh_time}
\begin{tabular}{cccccc} \ \\
\hline \hline
\multicolumn{3}{c}{Rank truncation ratio} & & \multicolumn{2}{c}{Time (s) for $H_\text{BSE}$}  \ \\
\hline
 $N_{vc}^{t}$/$N_{vc}$  &  $N_{vv}^{t}$/$N_{vv}$  & $N_{cc}^{t}$/$N_{cc}$ & & $\widetilde{V}_A$ & $\widetilde{W}_A$  \ \\
\hline
1.0 & 1.0 & 1.0 && 1.574 & 4.198 \ \\
1.0 & 0.5 & 0.1 && 1.574 & 1.474 \ \\
0.1 & 0.1 & 0.1 && 0.025 & 0.414 \ \\
\hline \hline
\end{tabular}
\end{table}

\subsection{Optical absorption spectra}\label{sec:Application}

\begin{figure}[!tb]
\centering
\includegraphics[width=0.95\textwidth]{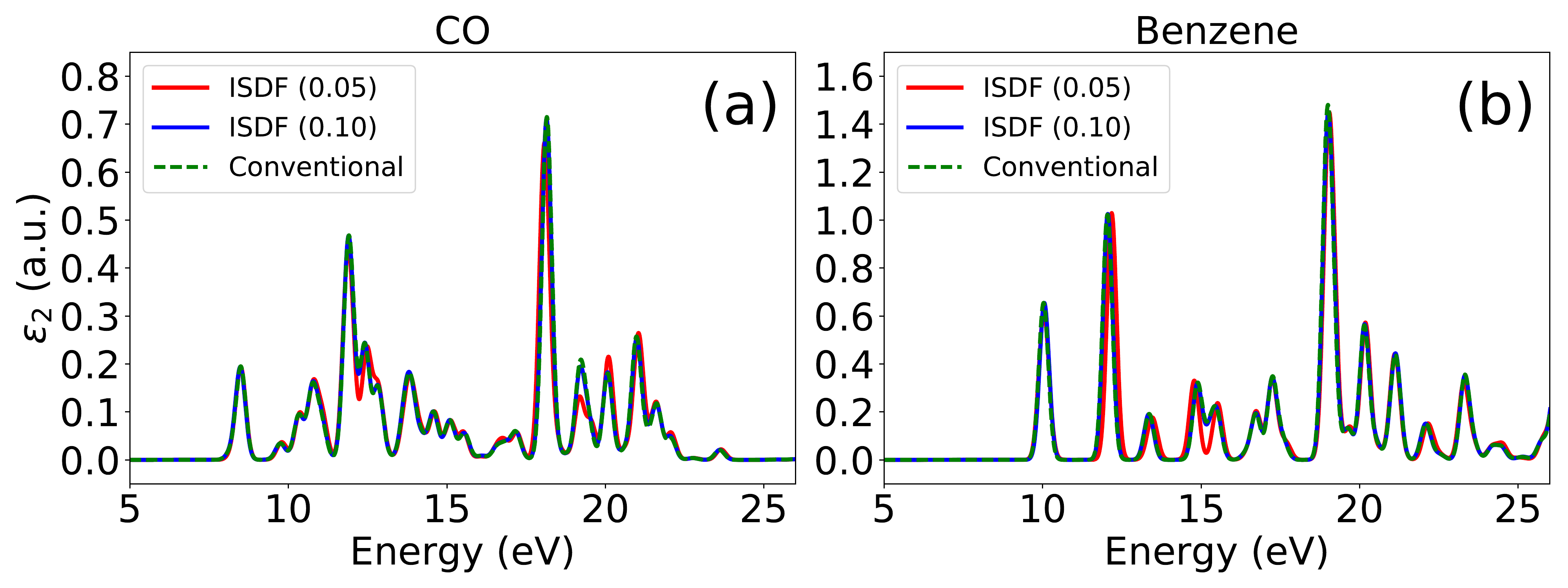}
\caption{Optical dielectric function (imaginary part
$\varepsilon_2$) of (a) CO and (b) benzene molecules computed with
the ISDF method (the rank ratio $N_{cc}^t$/$N_{cc}$ is set to be
0.05 ($t = 3.0$) and 0.10 ($t = 6.0$)) compared to conventional BSE
calculations in BerkeleyGW.} \label{fig:Absorption}
\end{figure}

One important application of BSE is to compute the optical
absorption spectrum, which is determined by optical dielectric
function in Equation~\eqref{eq:absorption}.
Fig.~\ref{fig:Absorption} plots the optical absorption spectra
for both CO and benzene obtained from approximate $H_\text{BSE}$
constructed with the ISDF method and the $H_\text{BSE}$
constructed in a conventional approach implemented in BerkeleyGW.
When the rank truncation ratio $N_{cc}^t$/$N_{cc}$
is set to be only 0.10 ($t = 6.0$), the absorption
spectrum obtained from the ISDF approximate $H_\text{BSE}$
is nearly indistinguishable from that produced from the conventional
approach.  When $N_{cc}^t$/$N_{cc}$ is set to
0.05 ($t = 3.0$), the absorption spectrum obtained from ISDF approximate $H_\text{BSE}$
still preserves the main features (peaks) of the absorption spectrum
obtained in a conventional approach even though some of the peaks are slightly shifted, and the height of some peaks are slightly off.

\section{Conclusion and outlook} \label{sec:Conclusion}

In summary, we have demonstrated that the interpolative separable density
fitting (ISDF) technique can be used to efficiently and accurately
construct the Bethe--Salpeter Hamiltonian matrix. The ISDF method
allows us to reduce the complexity of the Hamiltonian construction
from $\Or(N_e^5)$ to $\Or(N_e^3)$ with a much smaller pre-constant.
We show that the ISDF based BSE calculations in molecules and solids
can efficiently produce accurate exciton energies and optical
absorption spectrum in molecules and solids.

In the future, we plan to replace the costly QRCP procedure 
with the centroidal Voronoi tessellation (CVT)
method~\cite{arXiv_2017_CVT} for selecting the interpolation points in
the ISDF method. The CVT method is expected to significantly reduce the
computational cost for selecting interpolating point in the ISDF procedure for the BSE calculations.

The performance results reported here are based on a sequential
implementation of the ISDF method. In the near future, we
will implement a parallel version suitable for large-scale
distributed memory parallel computers. Such an implementation
will allow us to tackle much larger problems for which
the favorable scaling of the ISDF approach is much more
pronounced.


\section*{Acknowledgments}

This work is partly supported by the Center for Computational Study of
Excited-State Phenomena in Energy Materials (C2SEPEM) at the Lawrence Berkeley
National Laboratory, which is funded by the U.\,S. Department of Energy,
Office of Science, Basic Energy Sciences, Materials Sciences and Engineering
Division under Contract No. DE-AC02-05CH11231, as part of the Computational
Materials Sciences Program, and by the
Center for Applied Mathematics for Energy Research Applications
(CAMERA) (L.~L. and C.~Y.). The authors thank the National Energy
Research Scientific Computing (NERSC) center for making
computational resources available.


\begin{thebibliography}{10}

\bibitem{ScaLAPACK_2011}
T.~Auckenthaler, V.~Blum, H.~J. Bungartz, T.~Huckle, R.~Johanni, L.~Kr\"{a}mer,
  B.~Lang, H.~Lederer, and P.~R. Willems.
\newblock Parallel solution of partial symmetric eigenvalue problems from
  electronic structure calculations.
\newblock {\em Parallel Comput.}, 37(12):783--794, 2011.

\bibitem{JCP_334_221_2017}
P.~Benner, S.~Dolgov, V.~Khoromskaia, and B.~N. Khoromskij.
\newblock Fast iterative solution of the {Bethe}--{Salpeter} eigenvalue problem
  using low-rank and {QTT} tensor approximation.
\newblock {\em J. Comput. Phys.}, 334:221--239, 2017.

\bibitem{JCTC_11_5197_2015}
J.~Brabec, L.~Lin, M.~Shao, N.~Govind, Y.~Saad, C.~Yang, and E.~G. Ng.
\newblock Efficient algorithms for estimating the absorption spectrum within
  linear response {TDDFT}.
\newblock {\em J. Chem. Theory Comput.}, 11(11):5197--5208, 2015.

\bibitem{SIAM_13_727_1992_QRCP}
T.~F. Chan and P.~C. Hansen.
\newblock Some applications of the rank revealing {QR} factorization.
\newblock {\em SIAM J. Sci. Statist. Comput.}, 13:727--741, 1992.

\bibitem{CPC_183_1269_2012_BerkeleyGW}
J.~Deslippe, G.~Samsonidze, D.~A. Strubbe, M.~Jain, M.~L. Cohen, and S.~G.
  Louie.
\newblock {BerkeleyGW}: A massively parallel computer package for the
  calculation of the quasiparticle and optical properties of materials and
  nanostructures.
\newblock {\em Comput. Phys. Commun.}, 183(6):1269--1289, 2012.

\bibitem{arXiv_2017_CVT}
K.~Dong, W.~Hu, and L.~Lin.
\newblock Interpolative separable density fitting through centroidal voronoi
  tessellation with applications to hybrid functional electronic structure
  calculations.
\newblock {\em arXiv:1711.01531}, 2017.

\bibitem{JPCM_21_395502_2009_QE}
P.~Giannozzi, S.~Baroni, N.~Bonini, M.~Calandra, R.~Car, C.~Cavazzoni,
  D.~Ceresoli, G.~L. Chiarotti, M.~Cococcioni, I.~Dabo, A.~D. Corso,
  S.~de~Gironcoli, S.~Fabris, G.~Fratesi, R.~Gebauer, U.~Gerstmann,
  C.~Gougoussis, A.~Kokalj, M.~Lazzeri, L.~Martin-Samos, N.~Marzari, F.~Mauri,
  R.~Mazzarello, S.~Paolini, A.~Pasquarello, L.~Paulatto, C.~Sbraccia,
  S.~Scandolo, G.~Sclauzero, A.~P. Seitsonen, A.~Smogunov, P.~Umari, and R.~M.
  Wentzcovitch.
\newblock {QUANTUM ESPRESSO}: A modular and open-source software project for
  quantum simulations of materials.
\newblock {\em J. Phys.: Condens. Matter}, 21(39):395502, 2009.

\bibitem{PRB_54_1703_1996_LDA}
S.~Goedecker, M.~Teter, and J.~Hutter.
\newblock Separable dual-space {Gaussian} pseudopotentials.
\newblock {\em Phys. Rev. B}, 54:1703, 1996.

\bibitem{PRB_58_3641_1998_HGH}
C.~Hartwigsen, S.~Goedecker, and J.~Hutter.
\newblock Relativistic separable dual-space gaussian pseudopotentials from {H
  to Rn}.
\newblock {\em Phys. Rev. B}, 58:3641, 1998.

\bibitem{PR_139_A796_1965_GW}
L.~Hedin.
\newblock New method for calculating the one-particle {Green}'s function with
  application to the electron--gas problem.
\newblock {\em Phys. Rev.}, 139:A796, 1965.

\bibitem{JCTC_13_5420_2017_ISDF}
W.~Hu, L.~Lin, and C.~Yang.
\newblock Interpolative separable density fitting decomposition for
  accelerating hybrid density functional calculations with applications to
  defects in silicon.
\newblock {\em J. Chem. Theory Comput.}, 13(11):5420--5431, 2017.

\bibitem{MolPhys_114_1148_2016}
P.~B.~V. Khoromskaia and B.~N. Khoromskij.
\newblock A reduced basis approach for calculation of the {Bethe¨CSalpeter}
  excitation energies by using low-rank tensor factorisations.
\newblock {\em Mol. Phys.}, 114:1148--1161, 2016.

\bibitem{SIAMJSC_23_517_2001_LOBPCG}
A.~V. Knyazev.
\newblock Toward the optimal preconditioned eigensolver: Locally optimal block
  preconditioned conjugate gradient method.
\newblock {\em SIAM J. Sci. Comput.}, 23(2):517--541, 2001.

\bibitem{JRNBS_45_255_1950_Lanczos}
C.~Lanczos.
\newblock An iteration method for the solution of the eigenvalue problem of
  linear differential and integral operators.
\newblock {\em J. Res. Nat. Bur. Standards}, 45:255--282, 1950.

\bibitem{LinXuYing2017}
L.~Lin, Z.~Xu, and L.~Ying.
\newblock Adaptively compressed polarizability operator for accelerating large
  scale ab initio phonon calculations.
\newblock {\em Multiscale Model. Simul.}, 15:29--55, 2017.

\bibitem{PRB_92_075422_2015}
M.~P. Ljungberg, P.~Koval, F.~Ferrari, D.~Foerster, and
  D.~S{\'{a}}nchez-Portal.
\newblock Cubic-scaling iterative solution of the {Bethe}--{Salpeter} equation
  for finite systems.
\newblock {\em Phys. Rev. B}, 92:075422, 2015.

\bibitem{LuThicke2017}
J.~Lu and K.~Thicke.
\newblock {Cubic scaling algorithms for RPA correlation using interpolative
  separable density fitting}.
\newblock {\em J. Comput. Phys.}, 351:187--202, 2017.

\bibitem{JCP_302_329_2015}
J.~Lu and L.~Ying.
\newblock Compression of the electron repulsion integral tensor in tensor
  hypercontraction format with cubic scaling cost.
\newblock {\em J. Comput. Phys.}, 302:329--335, 2015.

\bibitem{PRB_95_075415_2017}
M.~Marsili, E.~Mosconi, F.~D. Angelis, and P.~Umari.
\newblock Large-scale {GW-BSE} calculations with {$N^3$} scaling: Excitonic
  effects in dye-sensitized solar cells.
\newblock {\em Phys. Rev. B}, 95:075415, 2017.

\bibitem{PR_46_618_1934_DMPT}
C.~M{\o}ler and M.~S. Plesset.
\newblock Note on an approximation treatment for many-electron systems.
\newblock {\em Phys. Rev.}, 46:618, 1934.

\bibitem{RMP_74_601_2002}
G.~Onida, L.~Reining, and A.~Rubio.
\newblock Electronic excitations: Density-functional versus many-body
  {Green}'s-function approaches.
\newblock {\em Rev. Mod. Phys.}, 74:601, 2002.

\bibitem{JCP_133_164109_2010}
D.~Rocca, D.~Lu, and G.~Galli.
\newblock Ab initio calculations of optical absorption spectra: Solution of the
  {Bethe¨CSalpeter} equation within density matrix perturbation theory.
\newblock {\em J. Chem. Phys.}, 133:164109, 2010.

\bibitem{PRB_62_4927_2000}
M.~Rohlfing and S.~G. Louie.
\newblock Electron--hole excitations and optical spectra from first principles.
\newblock {\em Phys. Rev. B}, 62:4927, 2000.

\bibitem{PR_84_1232_1951_BSE}
E.~E. Salpeter and H.~A. Bethe.
\newblock A relativistic equation for bound-state problems.
\newblock {\em Phys. Rev.}, 84:1232, 1951.

\bibitem{arXiv_1611.02348_2016}
M.~Shao, F.~H. da~Jornada, L.~Lin, C.~Yang, J.~Deslippe, and S.~G. Louie.
\newblock A structure preserving {Lanczos} algorithm for computing the optical
  absorption spectrum.
\newblock {\em SIAM J. Matrix.\ Anal.\ Appl.}, to appear.

\bibitem{LinearAlgebraAppl_488_148_2016}
M.~Shao, F.~H. da~Jornada, C.~Yang, J.~Deslippe, and S.~G. Louie.
\newblock Structure preserving parallel algorithms for solving the
  {Bethe¨CSalpeter} eigenvalue problem.
\newblock {\em Linear Algebra Appl.}, 488:148--167, 2016.

\bibitem{BSEPACK_UserGuide_2016}
M.~Shao and C.~Yang.
\newblock {BSEPACK} user's guide, 2016.
\newblock https://sites.google.com/a/lbl.gov/bsepack/.

\end{thebibliography}
\end{document}